\documentclass[10pt,a4paper,twoside]{article}
\usepackage{epsfig}
\usepackage{baltlat6}
\usepackage{array}
\usepackage{here}
\usepackage{lscape}
\renewcommand*\thefootnote{%
  \ifcase\value{footnote}\or
  *\or
  **\or
  ***\or
  ****\or
  *****\or
  ******\or
  *******\or
  ********\or
  *********\fi
}

\newcommand{\sla}{\raisebox{-0.10em}{$\stackrel{<}{{\mbox{\tiny $\sim$}}}$}}

\begin{document}
\ \
\vspace{0.5mm}
\setcounter{page}{1}
\vspace{8mm}

\titlehead{Baltic Astronomy}

\titleb{SPECTRAL ANALYSIS VIA THE VIRTUAL OBSERVATORY:\\ THE SERVICE THEOSSA}

\begin{authorl}
\authorb{E. Ringat}{},
\authorb{T. Rauch}{} and
\authorb{K. Werner}{}
\end{authorl}

\begin{addressl}
\addressb{}{Institute for Astronomy and Astrophysics, Kepler Center for Astro and Particle Physics, Eberhard Karls University, Sand 1, 72076 T\"ubingen, Germany;
gavo@listserv.uni-tuebingen.de}
\end{addressl}

\submitb{Received: 2011; accepted: 2011 }

\begin{summary} In the framework of the Virtual Observatory, the newly developed service \emph{TheoSSA} provides access to theoretical stellar spectral-energy distributions. In a pilot phase, this service is based on the well established T\"ubingen NLTE Model-Atmosphere Package for hot, compact stars. We demonstrate its present capabilities and future extensions.
\end{summary}

\begin{keywords} Methods: data analysis -- Stars: atmospheres -- Stars:  fundamental
parameters \end{keywords}

\resthead{SPECTRAL ANALYSIS VIA THE VO: THEOSSA}
{E. Ringat, T. Rauch, K. Werner}

\sectionb{1}{INTRODUCTION}

Spectral-analysis techniques and stellar model-atmosphere codes improved over the last 40 years. Beginning with LTE codes and mainly hydrogen models, we arrived today at sophisticated, fully metal-line blanketed NLTE model-atmospheres, considering all elements from hydrogen to nickel.

One of these codes is the T\"ubingen NLTE Model-Atmosphere Package (\emph{TMAP}\footnote{http://astro.uni-tuebingen.de/\raisebox{.3em}{\tiny $\sim $}TMAP}, Werner et al.\@ 2003, Rauch \& Deetjen 2003) that has its roots in the 1980s (Werner 1986). The related atomic database \emph{TMAD}\footnote{http://astro.uni-tuebingen.de/\raisebox{.3em}{\tiny $\sim $}TMAD} (T\"ubingen Model-Atom Database) was built up together with the code and is steadily updated. \emph{TMAP} assumes hydrostatic and radiative equilibrium. It is suitable for the analysis of hot, compact objects with $20\,000\,$K$ \ \sla\ \ T_\mathrm{eff} \ \sla\  200\,000$\,K and $4 \ \sla\  \log g \ \sla\  9$. Up to 1500 NLTE levels and about 4000 lines for hydrogen up to potassium, and hundreds of millions of lines for the iron-group elements (Ca$-$Ni) can be considered.

Within the framework of the \,German \,Astrophysical \,Virtual \,Observatory  (\emph{GAVO}\footnote{http://g-vo.org/}), the German share in the International Virtual Observatory Alliance (\emph{IVOA}), \emph{TMAP} and \emph{TMAD} were made accessible to the community. In three ways, spectral-energy distributions (SEDs) can be accessed. These ways are described in the following sections.

\begin{landscape}
\begin{figure}
\vbox{
\centerline{\psfig{figure=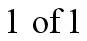,height=100mm,angle=0,clip=}}
\vspace{1mm}
\captionb{1}
{Web interface of \emph{TheoSSA}.}
}
\end{figure}
\end{landscape}

\begin{landscape}
\begin{figure}[!tH]
\vbox{
\centerline{\psfig{figure=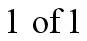,height=120mm,angle=0,clip=}}
\vspace{1mm}
\captionb{2}
{Results of a \emph{TheoSSA} query (Fig.\,1).}
}
\end{figure}
\end{landscape}
\begin{landscape}
\begin{figure}[!tH]
\vbox{
\centerline{\psfig{figure=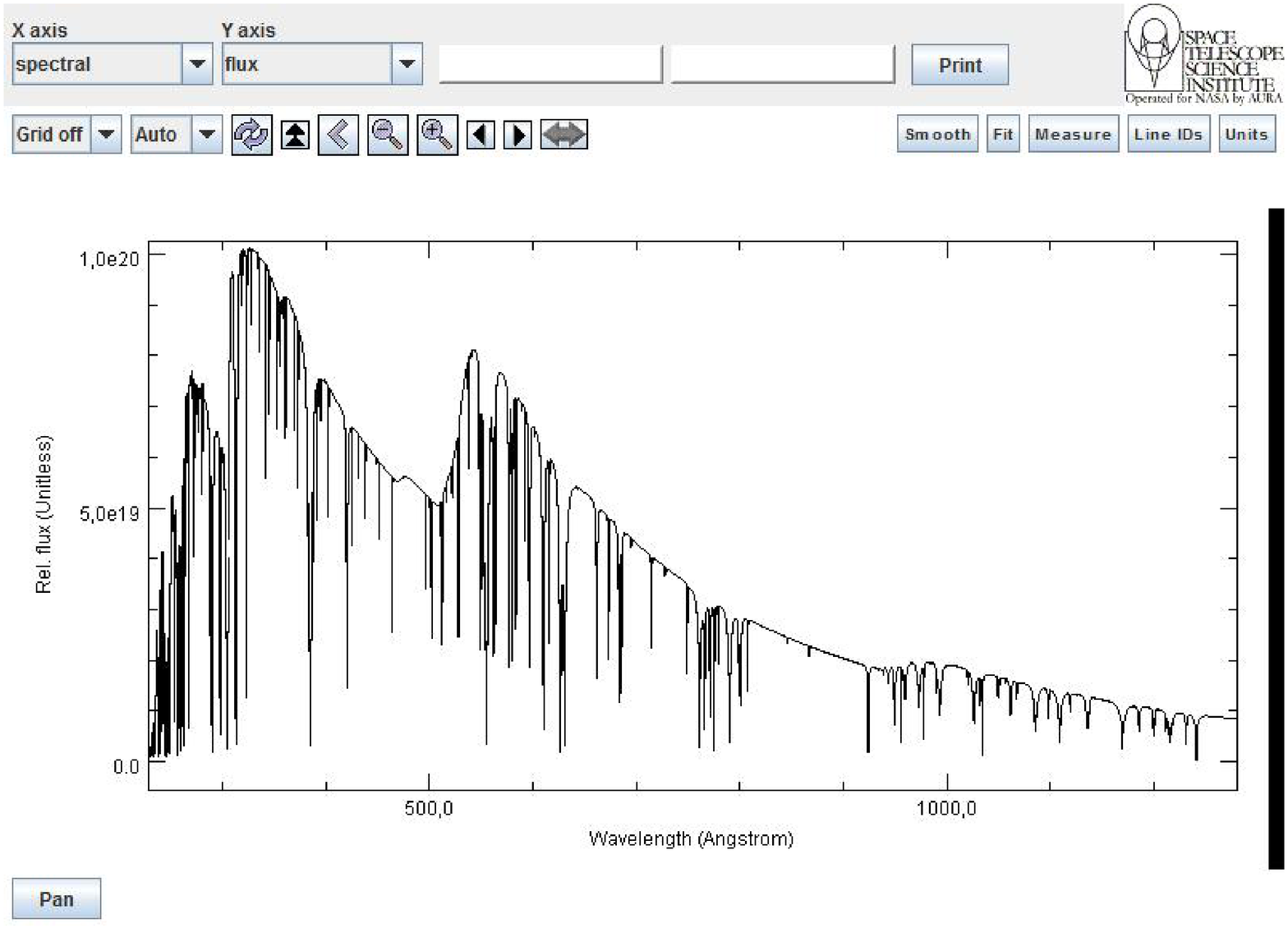,height=120mm,angle=0,clip=}}
\vspace{1mm}
\captionb{3}
{A \emph{TheoSSA} spectrum displayed with \emph{SPECVIEW}.}
}
\end{figure}

\end{landscape}

\sectionb{2}{DOWNLOADING PRE-CALCULATED SEDS - \emph{TheoSSA}}

\emph{TheoSSA}\footnote{http://dc.g-vo.de/theossa} provides access to a large database containing stellar SEDs. It uses VO standards and is designed to contain stellar spectra of all model-atmosphere codes in VO compliant format.

\begin{figure}[!tH]
\vbox{
\centerline{\psfig{figure=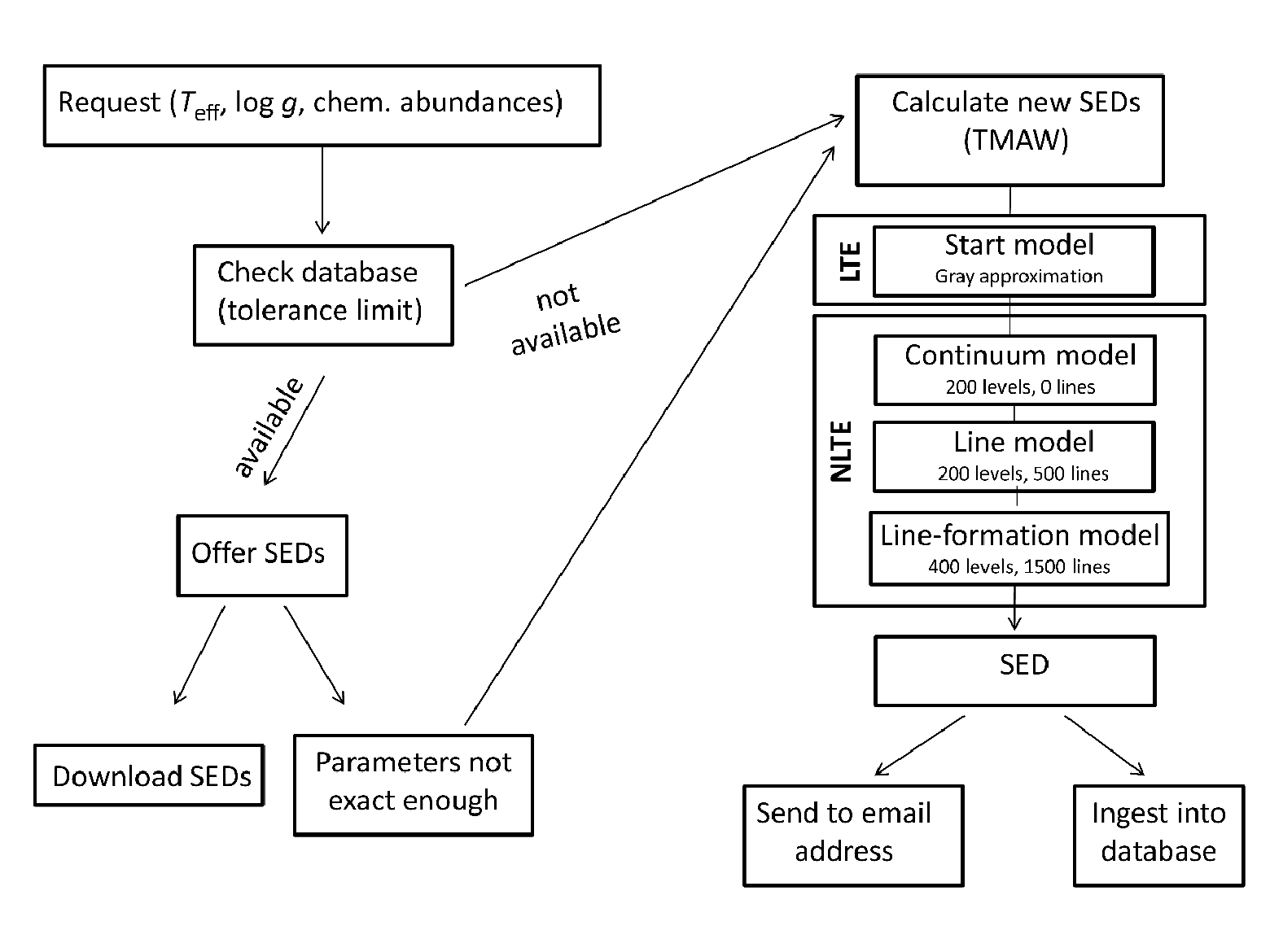,width=\textwidth,angle=0,clip=}}
\vspace{1mm}
\captionb{4}
{Scheme of \emph{TheoSSA} and \emph{TMAW} with typical level and line numbers.}
}
\end{figure}

With its web interface (Figure\,1), the usage of \emph{TheoSSA} is very fast and easy. The desired parameters ($T_\mathrm{eff}$, $\log g$, and abundances) are entered and a list of available SEDs within a given parameter range is returned (Figure\,2). For a quick overview, the SEDs can be plotted with \emph{SPECVIEW}\footnote{http://www.stsci.edu/resources/software$\_$hardware/specview} (Figure\,3). The spectra can be downloaded in VOTable or plain text format, where the metadata of the spectra is written in the header part.

If the requested SED is not in the database, it can be calculated with the service \emph{TMAW} (Figure\,4). All newly calculated SEDs are automatically ingested into the \emph{TheoSSA} database and, thus, it is growing in time. The \emph{TMAW} SEDs are identified in their meta data by the entry ``DataID.Creator=TMAW''.

\newpage
\sectionb{3}{CALCULATING DETAILED SEDS - \emph{TMAW}} 

With the web interface of the model-atmosphere code (\emph{TMAW}\footnote{http://astro.uni-tuebingen.de/\raisebox{.3em}{\tiny $\sim$}TMAW}), individual SEDs containing hydrogen, helium, carbon, nitrogen, and oxygen can be requested. The service is also controlled via web interface, where the fundamental parameters as well as information about the requester are entered. \emph{TMAW} can be used as a blackbox, because no detailed knowledge of the code is necessary. However, to do reliable science with this service, having some background knowledge is important.

\emph{TMAW} calculates a model generally in three steps (Figure\,4). Firstly a start model is created. Then a continuum model is calculated, starting with a fixed temperature structure, then using the Uns\"old-Lucy temperature correction (Uns\"old 1968). Subsequently line opacities are considered, again beginning without temperature correction and using the Uns\"old-Lucy correction later. Finally the SED is calculated with more detailed model atoms (including fine-structure splitting).

Presently, \emph{TMAD} contains data for the elements H, He, C, N, O, F, Ne, Na, Mg, Si, S, Ar, and Ca. As an example, in Figure\,5 the \emph{TMAD} O\,\textsc{v} model atom is displayed. Because \emph{TMAD} was originally built for \emph{TMAP}, its format is compatible with that code. It can be used for every model-atmosphere code. \emph{TMAD} is continuously updated.

The database contains two kinds of data. For stellar atmosphere calculations, ready-to-use model atoms for distinct temperature ranges can be downloaded or individual model atoms can be tailored. For SED calculations the files contain fine-structure splitting.
\begin{figure}[!tH]
\vbox{
\centerline{\psfig{figure=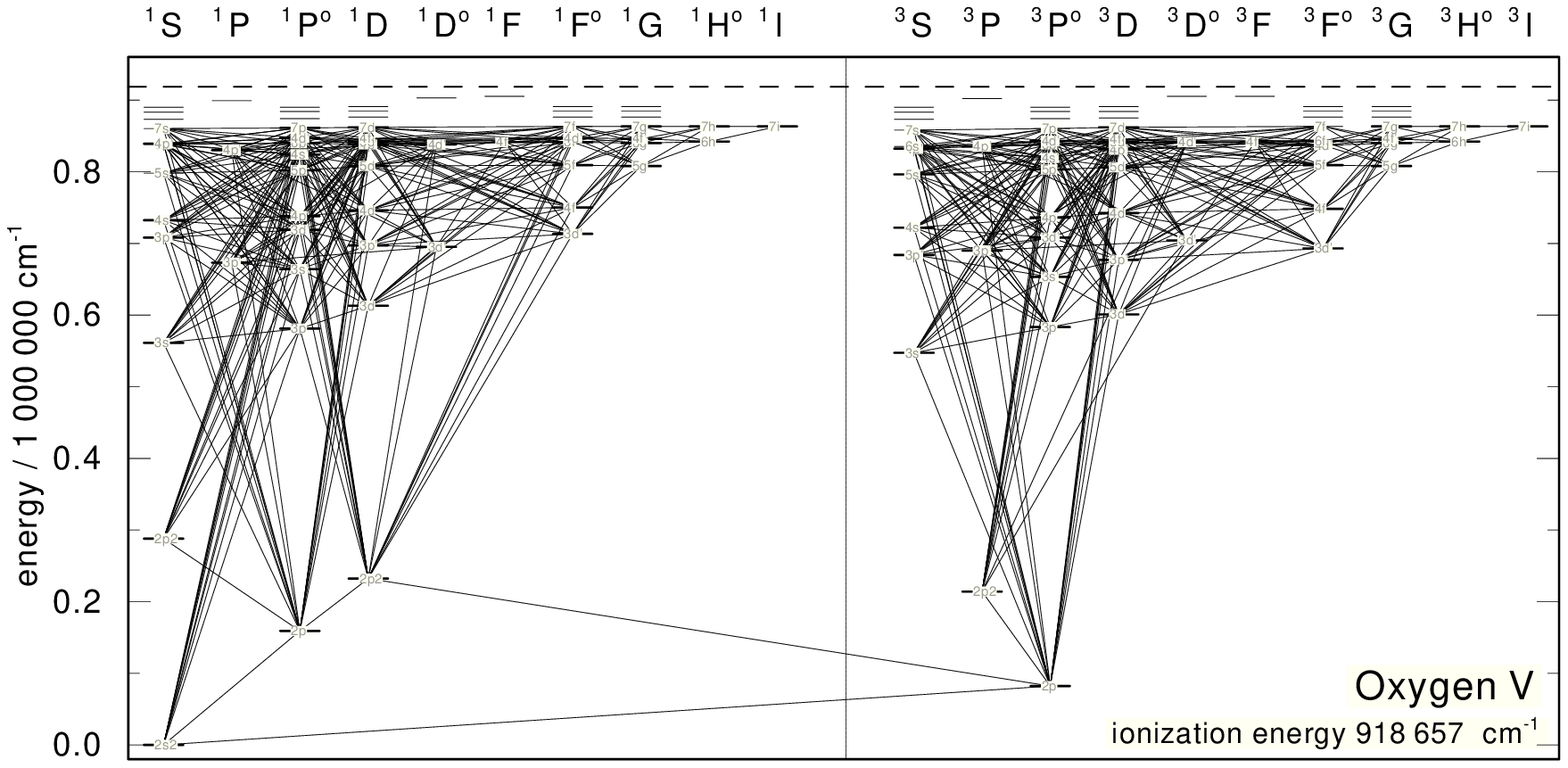,width=\textwidth,angle=0,clip=}}
\vspace{1mm}
\captionb{5}
{Grotrian diagram of O\,\textsc{v} (\emph{TMAD}).}
}
\end{figure}

\sectionb{4}{RECENT IMPROVEMENTS}

Originally \emph{TMAW} was designed to create SEDs as input for photoionization models for planetary nebula (PN) simulations (e.g.\@ Ercolano et al.\@ 2003). In the last century, many PN analyses used blackbodies to simulate the stellar, ionizing flux. Blackbodies vastly differ from models calculated with a model-atmosphere code like \emph{TMAP} (Figure~6). For this purpose, \emph{TMAW} and \emph{TheoSSA} supply SEDs in a fast and easy way. 

To ensure a reasonable calculation time for \emph{TMAW}, relatively small model atoms and the numerically stable Uns\"old-Lucy temperature correction are used, and opacities for H, He, C, N, and O, only, are considered. This raises the question how reliable are \emph{TMAW} models for spectral analyses? The aim of \emph{TMAW} is to achieve an accuracy $<$\,20\% in the determination of basic photospheric parameters like $T_\mathrm{eff}$, $\log g$, and elemental abundances. We wanted to check if this is enough for reliable spectral analysis. Therefore we re-calculated a grid of 527 models with \emph{TMAW}, originally calculated with \emph{TMAP} (including opacities of many species from H to Ni) for the analysis of the sdOB star AA Doradus (Klepp \& Rauch 2011). We performed a $\chi^2$ fit analogously to them, and found only small deviations of about 1\,\% in $T_\mathrm{eff}$ and 7\,\% in $\log g$ (Ringat 2011). Thus, \emph{TMAW} SEDs appear well suited for the analysis of optical spectra. 

\begin{figure}[!tH]
\vbox{
\centerline{\psfig{figure=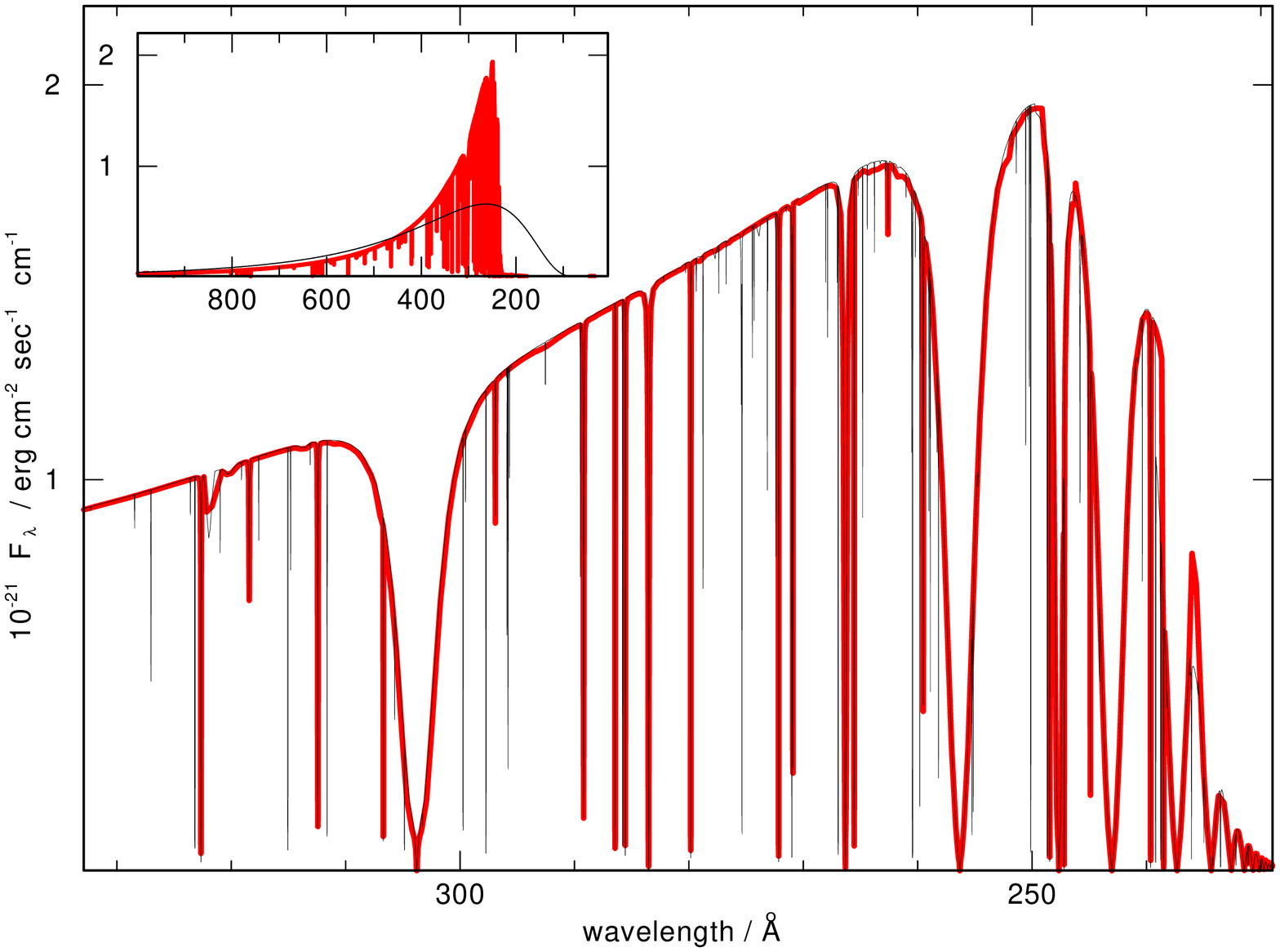,angle=-90,width=\textwidth,clip=}}
\vspace{1mm}
\captionb{6}
{\emph{TMAP} (thin line) flux ($T_\mathrm{eff}=110\,000$\,K, $\log g=7.5$) compared with a respective \emph{TMAW} flux (thick, red line in the online version). The insert shows the strong deviation of a blackbody flux (with respective temperature) from a model-atmosphere spectrum.}
}
\end{figure}

In order to further improve \emph{TMAW}, we included now additional iteration steps with the standard \emph{TMAP} temperature correction and a subsequent line-formation calculation with far more detailed model atoms. With this method the accuracy is improved without strongly increasing the calculation time.

\clearpage
\sectionb{5}{SUMMARY}

Within the framework of the \emph{GAVO} project, spectral energy distributions calculated with the T\"ubingen NLTE Model-Atmosphere Package are available in three ways:
\begin{itemize}
\item pre-calculated SEDs can be downloaded via \emph{TheoSSA}
\item missing / more individual SEDs can be calculated via a \emph{TMAW} request
\item model atoms can be downloaded and tailored and used for every model-atmosphere code
\end{itemize}

These services supply the VO user with sophisticated SEDs in a very fast and easy way. They are suitable as input for photoionization models and for spectral analyses. To increase the accuracy, we are implementing post-calculation steps that use the temperature correction used by \emph{TMAP} and more detailed model atoms.

\sectionb{ACKNOWLEDGEMENTS}{}
 ER is supported by the Federal Ministry of  Education and Research (BMBF) under grant 05A11VTB, TR by the German Aerospace Center (DLR) under grant 05OR 0806. The \emph{TheoSSA} service (http://dc.g-vo.org/theossa) used to retrieve theoretical spectra for this
    paper was constructed as part of the activities of the German Astrophysical Virtual Observatory.

\References
\refb Ercolano,~B., Barlow,~M.~J., Storey,~P.~J. et al.\@ 2003, MNRAS, 344, 1145
\refb Klepp,~S. \& Rauch,~T. 2011, A\&A, 531, L7
\refb {Rauch},~T. \& {Deetjen}, J.\@~L.\@ 2003,
   in: I.~Hubeny, D.~Mihalas, \& K.~Werner (eds.),
   \textit{Stellar Atmosphere Modeling},
         ASP-CS, 288, 103 
\refb {Ringat}, E. 2011,
	in: D.~Kilkenny, S.~Jeffery, \& C.~Koen (eds.),
		\textit{The Fifth Meeting on Hot Subdwarf Stars \& Related Objects}, in press (arXiv:1111.0546v1)
\refb Uns\"old,~A. 1968,
	\textit{Physik der Sternatmosph\"aren}, Springer (Berlin), 2nd edition
\refb Werner, K. 1986, A\&A, 161, 177 
\refb {Werner}, K., {Deetjen}, J.~L., {Dreizler}, S., et al.\@  2003,
    in: I.~Hubeny, D.~Mihalas, \& K.~Werner (eds.),
    \textit{Stellar Atmosphere Modeling},
         ASP-CS, 288, 31

\end{document}